\documentclass[10pt,journal]{IEEEtran}

\usepackage{graphicx,amssymb}
\graphicspath{{figures/}}
\usepackage{float}
\usepackage{wrapfig}
\usepackage{fancyhdr}    
\usepackage{fancybox}
\usepackage{amsmath}
\usepackage{latexsym,amsmath,url,caption,epsfig,cite,array}
\usepackage{circuitikz}
\usepackage{mathrsfs}   
\usepackage{verbatim}
\usepackage{pgfplots} 
\usetikzlibrary{patterns,pgfplots.fillbetween}
\pgfplotsset{compat=1.3, tick style={black}}
\usepackage{multirow}
\usepackage{xcolor}
\usepackage[font=footnotesize,caption=false,subrefformat=parens,labelformat=parens]{subfig}
\usepackage{algorithm}
\usepackage{algorithmic}
\usepackage[english]{babel}
\usepackage[utf8]{inputenc}
\usepackage{cutwin,lipsum}
\usepackage{soul}
\usepackage{mathtools}
\usepackage{nomencl}
\makenomenclature
\usepackage{amsmath,amssymb,amsfonts,bm}

\usepackage{hyperref}
\hypersetup{
    colorlinks=true,
    linkcolor=blue,     
    urlcolor=cyan,
    }

\usepackage{mathtools}
\DeclarePairedDelimiter\ceil{\lceil}{\rceil}

\title{Data-driven Distributed Control to Scale EV Integration into Power Grid}

 \author{\IEEEauthorblockN{Emin Ucer and Mithat Kisacikoglu}
 \\
 \IEEEauthorblockA{Dept. of Electrical and Computer Engineering,
 The University of Alabama, Tuscaloosa, AL\\
 Emails: eucer@crimson.ua.edu and mkisacik@ua.edu }

 \thanks{This material is based upon work supported by the National Science Foundation under Award No~1755996.}
 }

\begin{document}

\IEEEoverridecommandlockouts

\IEEEpubid{\begin{minipage}[t]{\textwidth}\ \\[10pt]
        \centering\footnotesize
        {~\copyright2022 IEEE Personal use of this material is permitted.  Permission from IEEE must be obtained for all other uses, in any current or future media, including reprinting/republishing this material for advertising or promotional purposes, creating new collective works, for resale or redistribution to servers or lists, or reuse of any copyrighted component of this work in other works.}
\end{minipage}} 

\maketitle

\begin{abstract}

Electric vehicles (EVs) are finally making their way onto the roads, but the challenges concerning long charging times and impact on congestion of the power distribution grid are still not resolved. Proposed solutions depend on heavy communication and rigorous computation and mostly need real-time connectivity for optimal operation; thereby, they are not scalable. With the availability of historical measurement data, EV chargers can take better-informed actions while staying mostly off-line. This study develops a distributed and data-driven congestion detection methodology together with the Additive Increase Multiplicative Decrease (AIMD) algorithm to control mass EV charging in a distribution grid. The proposed distributed AIMD algorithm performs very closely to the ideal AIMD in terms of fairness and congestion handling, and its communication need is significantly low. The results can provide crucial insights on how data can be used to reveal the inner dynamics and structure of the power grid and help develop more advanced data-driven algorithms for grid integrated power electronics control.

\vspace{-4mm}

\end{abstract}
\section{Introduction}
High penetration of electric vehicles (EVs) with uncontrolled charging will cause transformer and line congestion in the power distribution grid. Among some adverse effects of this congestion are severe voltage drops, increased peak loading, thermal overheating, and even failure of equipment~\cite{erden2015examination,fernandez2011assessment,shafiee2013investigating,veldman2015distribution,leemput2014impact}. Therefore, control of EV charging has become an important research effort to mitigate these issues. Demand-side load management is a technique that is used to modify customer demand through various tools and methods. Curtailing this demand at peak hours is known as \textit{peak-shaving}. It is used to eliminate short-term demand spikes by smoothing out peak loads, preventing equipment overloading. This study will investigate a peak-shaving methodology to reduce the substation peak loading caused by mass EV integration.

Conventional methods for EV charging control require excessive system information, e.g., the grid topology, load forecasting, and customer preferences~\cite{Richardson2012Optimal,Sojoudi2011Optimal,Restrepo2018Three,Zhang2017Scalable,Liu2017Electric}. Uninterrupted connectivity is also needed to communicate this information to the EV chargers and send/receive control commands to/from the chargers. Assuming these are available, various optimization problems have been formulated to achieve certain objectives (e.g. minimizing generation cost, losses, and peak load; maximizing capacity utilization and total charging power) while respecting system constraints such as voltage limits and equipment overloading.

A traditional approach is to form an optimal power flow (OPF) problem and deterministically solve it using forecasted generation and demand values to determine the control setpoints. Recently, machine learning (ML) has also been used to develop control algorithms that can evolve based on OPF simulations~\cite{Ibrahim2020Machine}. The idea is to develop local controller rules from a dataset generated by OPF solutions. In general, performing these optimization tasks is challenging since the required system model and inputs are either missing or very hard to obtain completely. In \cite{Karagiannopoulos2019Datadriven} and \cite{Dobbe2020Toward}, data-driven local controllers are proposed by solving an off-line OPF problem using ML regression models. Since these controllers are trained based on a historical dataset, they are vulnerable to novel scenarious as well as instances caused by policy and natural shifts~\cite{Dobbe2020Learning}.

On the other hand, model-free approaches are generic and not dependant on the OPF model, i.e., system topology and variables. They usually use measurements (voltage, frequency, etc.) or historical data to obtain information regarding the grid's dynamic behavior. In this line of work, we presented a decentralized EV charging solution that uses the local voltage measurements and adapts the Internet's TCP/IP protocol for congestion detection~\cite{Ucer2020Decentralized}. 
This protocol implements the Additive Increase Multiplicative Decrease (AIMD) algorithm based on local measurements by implicitly detecting the congestion event (CE) without requiring any explicit feedback signal or system knowledge. In the context of distribution grid, the CE corresponds to the overloading of substation transformer feeder that powers the distribution network. Note that the CE can also be triggered when there is a need to shave the peak loading in a demand side load management scenario. Our previous works \cite{Ucer2020Decentralized,ucer2018analysis,ucer2019internet,ucer2019analysisPESGM} 
investigated implicit ways of estimating CE in the distribution grid and proposed algorithms that work with only local voltage measurements. However, the previously proposed algorithms can detect a fictitious congestion if their parameters are poorly tuned. True congestion detection requires proper tuning of the parameters, which is only possible via heuristics or by simulating the system if topology and load profiles are known. Thus, they can primarily operate sub-optimal. For this reason, we developed and presented a data-driven feeder loading estimation method in \cite{Ucer2020Machine} that only requires historical data, and we verified the method with real field measurements. As a next step, this work further develops and integrates the data-driven loading estimation idea presented in \cite{Ucer2020Machine} into our AIMD based EV charging algorithm, and implements it in a realistically modeled distribution grid simulation to asses its performance and effectiveness.

In this study, we explore how local historical measurement data can be utilized to learn the grid's dynamic status and how AIMD-based EV charging control can operate through this learned behavior. In particular, we propose to learn the relation between local measurements (i.e., voltage and phase) and the substation feeder loading using ML on historical data so that EVs can better estimate when CE occurs by observing their local voltage. Our methodology is to develop an ML model that maps local voltage to feeder loading and use this model as feedback for the AIMD controller. This will help us estimate the substation load in real-time using local voltage measurements without requiring connectivity to a central controller. Hence, our approach differs from other studies because we do not require system topology and loading information for operation (model-free), and we do not need a real-time feedback signal (reduced communication). We also do not develop rules for optimum controller actions using ML.
\vspace{-0.1cm}


\section{Analysis of End-node Voltage vs. Feeder Power Relationship in a Radial Distribution Grid} \label{sec:grid_analysis}

In the ideal AIMD algorithm, end nodes need to be notified by the CE. This could be performed by a direct and uninterrupted communication network, making it vulnerable during communication failures. It is much less demanding in terms of communication needs if the local end-nodes can detect grid congestion. In this regard, local voltage measurements can be an indicator of congestion information. The voltage varies depending on the loading level on the grid. Therefore, each node can make an estimation of the substation loading by observing their voltage if they have a guideline that maps their voltage level to the total power drawn from the substation. This guideline can be referred to as the voltage and power relationship~\cite{Ganu2013nPlug,Ucer2020Machine}.
\begin{figure}[t]
\centering
\includegraphics[trim=1.1cm 0.6cm 1.1cm 0.6cm, scale=0.22]{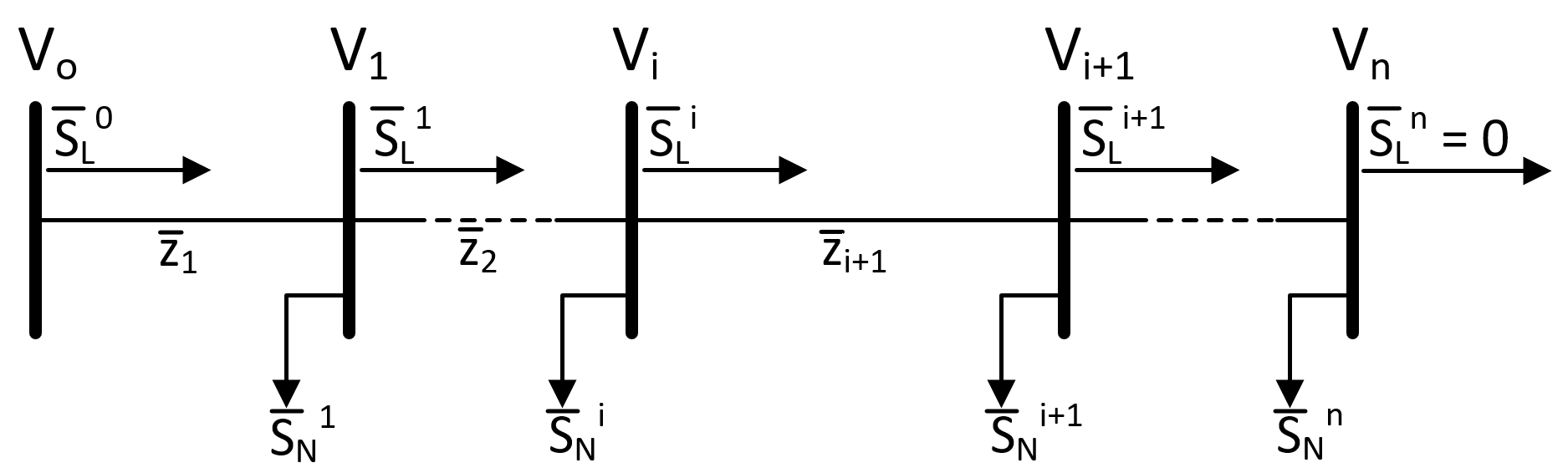}
\caption{Single feeder, radial distribution grid model.}
\label{fig:grid_model}
\end{figure}

We demonstrate this relationship between node voltage and total demand power on a single feeder, radial distribution grid model shown in Fig.~\ref{fig:grid_model}. In this model, $V_{o}$ represents the substation voltage magnitude, $\{V_{1},...,V_{n}\}$ are magnitudes of node voltages at nodes $1,\ldots, n$, and $\overline{S_L^i} = {P_L^i} + j{Q_L^i}$ (subscript $L$ stands for line) denotes the complex power flowing from node $i$ to node $i+1$ over a line impedance of $\overline{Z_{i}} = r_{i} + jx_{i}$. $\overline{S_N^i} = {P_N^i} + j{Q_N^i}$ (subscript $N$ stands for node) is the complex power drawn from node $i$. This model can be solved for an end-node voltage $V_{i}$ using the distribution grid branch flow equations (\textit{DistFlow} \cite{Baran1989Optimal}) as shown in \eqref{eq:DistFlow}.

By further simplifying the equations using the \textit{LinDistFlow} approximation~\cite{Baran1989Optimal}, we can derive an expression for the voltage of $i^{th}$ node as in \eqref{eq:VoltageVsPower}. The equation \eqref{eq:VoltageVsPower} shows that the end-node voltage ($V_{i}$) can be expressed as a combination of total and individual active and reactive power components scaled by the topology parameters. The goal is to learn the parameters of a function that maps the end-node voltage ($V_{i}$) to substation power such that $f(V_{i}) \to P_{L}^{0}, Q_{L}^{0}$.
\begin{equation} \label{eq:DistFlow}
\begin{split}
    {P_L^{i+1}} &{=}{P_L^i}{-} r_{i+1}\frac{{P_L^i}^{2}{+}{Q_L^i}^{2}}{V_{i}^{2}}{-}{P_N^{i+1}}, \\
    {Q_L^{i+1}} &{=}{Q_L^i}{-} x_{i+1}\frac{{P_L^i}^{2}{+}{Q_L^i}^{2}}{V_{i}^{2}}{-}{Q_N^{i+1}}, \\
     V_{i+1}^{2} &{=}V_{i}^{2}{-}2(r_{i+1}{P_L^i}{+} x_{i+1}{Q_L^i}){+}(r_{i+1}^{2}{+} x_{i+1}^{2})\frac{{P_L^i}^{2}{+}{Q_L^i}^{2}}{V_{i}^{2}}.
\end{split}
\end{equation}
 Utilizing the historical local voltage data ($V_{i}$) and total feeder power data ($P_{L}^{0}$ and $Q_{L}^{0}$), the problem of estimating grid loading level using local voltage information turns into a supervised learning problem. One reasonable simplification can be made based on the fact that total reactive power consumption $Q_{L}^{0}$ in a distribution power grid is relatively much smaller relative to total active power consumption $P_{L}^{0}$. Thus, total apparent power $S_{L}^{0}{=}|\overline{S_L^0}|$ can be assumed to be very close to $P_{L}^{0}$. Therefore, we can construct the mapping from $V_{i}$ to $S_{L}^{0}$ such that $f(V_{i}) \to S_{L}^{0}$. This assumption reduces the mapping output to a single variable ($S_{L}^{0}$).
\begin{equation} \label{eq:VoltageVsPower}
\begin{split}
V_{i}^{2} & =V_{0}^{2}-2(\overbrace{P_{L}^{0}}^{\substack{\text{Substation} \\ \text{Total P}}} \underbrace{\sum_{1}^{i} r_{i}}_\text{Constant}{}+\overbrace{Q_{L}^{0}}^{\substack{\text{Substation} \\ \text{Total Q}}} \underbrace{\sum_{1}^{i} x_{i}}_\text{Constant}) \\
& + 2\underbrace{[P_{N}^{1} P_{N}^{2} \cdots P_{N}^{i-1}]}_\text{End-node Active Powers}\underbrace{\begin{bmatrix} r_{2}+r_{3}+\cdots+r_{i}   \\ r_{3}+\cdots+r_{i}  \\ \vdots \\r_{i}   \end{bmatrix}}_\text{Constant} \\
& + 2\underbrace{[Q_{N}^{1} Q_{N}^{2} \cdots Q_{N}^{i-1}]}_\text{End-node Reactive Powers}\underbrace{\begin{bmatrix} x_{2}+x_{3}+\cdots+x_{i}   \\ x_{3}+\cdots+x_{i}  \\ \vdots \\x_{i}   \end{bmatrix}}_\text{Constant} \\
\end{split}
\end{equation}

However, several factors affect this mapping. First, $V_{0}$ will have some variations throughout the day 
even though it is usually highly regulated at the feeder level. Second, voltage regulation devices such as on-load tap changers (OLTCs), voltage regulators (VRs), and capacitor banks operate at different points in the network. They can change the end-node voltage and thus affect the relationship. Third, distribution grid topology can change due to various reasons, including system reconfiguration and expansion. Fourth, end-node active and reactive power ($ [P_{N}^{1} P_{N}^{2} \cdots P_{N}^{i-1}]$ and $[Q_{N}^{1} Q_{N}^{2} \cdots Q_{N}^{i-1}]$) are stochastic parameters that change in time creating a noise effect for the mapping model.
Last, significant reactive power consumption and generation will also deteriorate the relationship impacting the parameters.

All these factors make the problem very challenging and therefore require rigorous analysis. To eliminate the effects of some of these factors and improve the accuracy of mapping, we introduced two more feature variables, namely phase angle ($\delta_{i}$) and time interval ($T_{k}$). The phase angle of end-nodes changes in proportion to the total power due to the reactance of distributions lines. Also, splitting time into different intervals can help identify time dependencies of total power consumption with the aid of ML. 
\section{AIMD-Based EV Charging Algorithm} \label{sec:algorithm}
By providing historical substation loading information to end-users, every user can learn the relationship between its local variables and substation loading by locally building an ML regression model. For this reason, the historical substation loading data is made to be accessible to every end-node via a communication network (e.g., the Internet), and end-nodes can download this data only when they need to train their ML model. We note that this is not a real-time feedback signal for our controller. Instead, it is historical power measurements collected at the substation feeder level.

\begin{figure}[b]
\centering
\includegraphics[trim=0.5cm 0.3cm 0.5cm 0.3cm, scale=0.56]{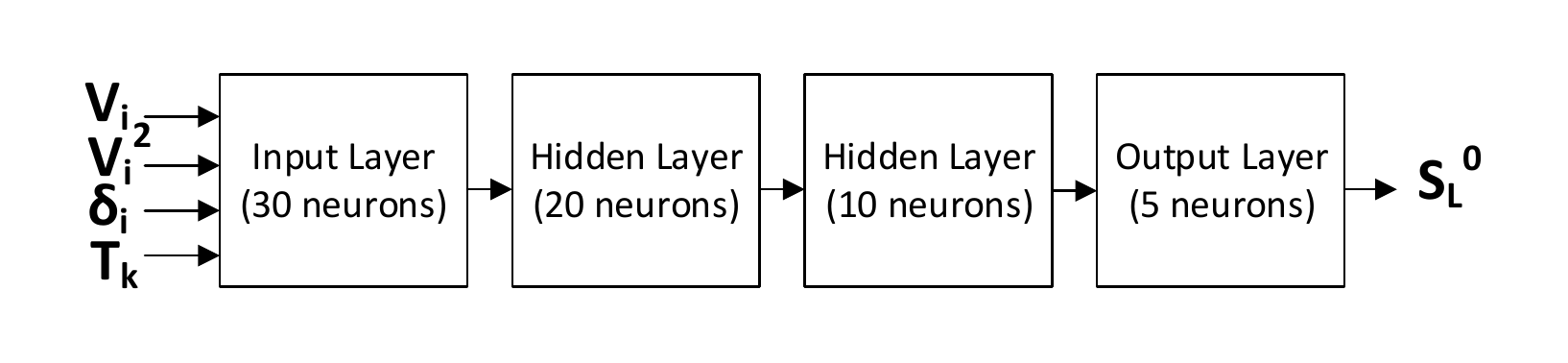}
\vspace{-0.5cm}
\caption{Neural network structure.}
\label{fig:neural_network}
\end{figure}

\begin{figure}[b]
\centering
\includegraphics[scale=0.195 ]{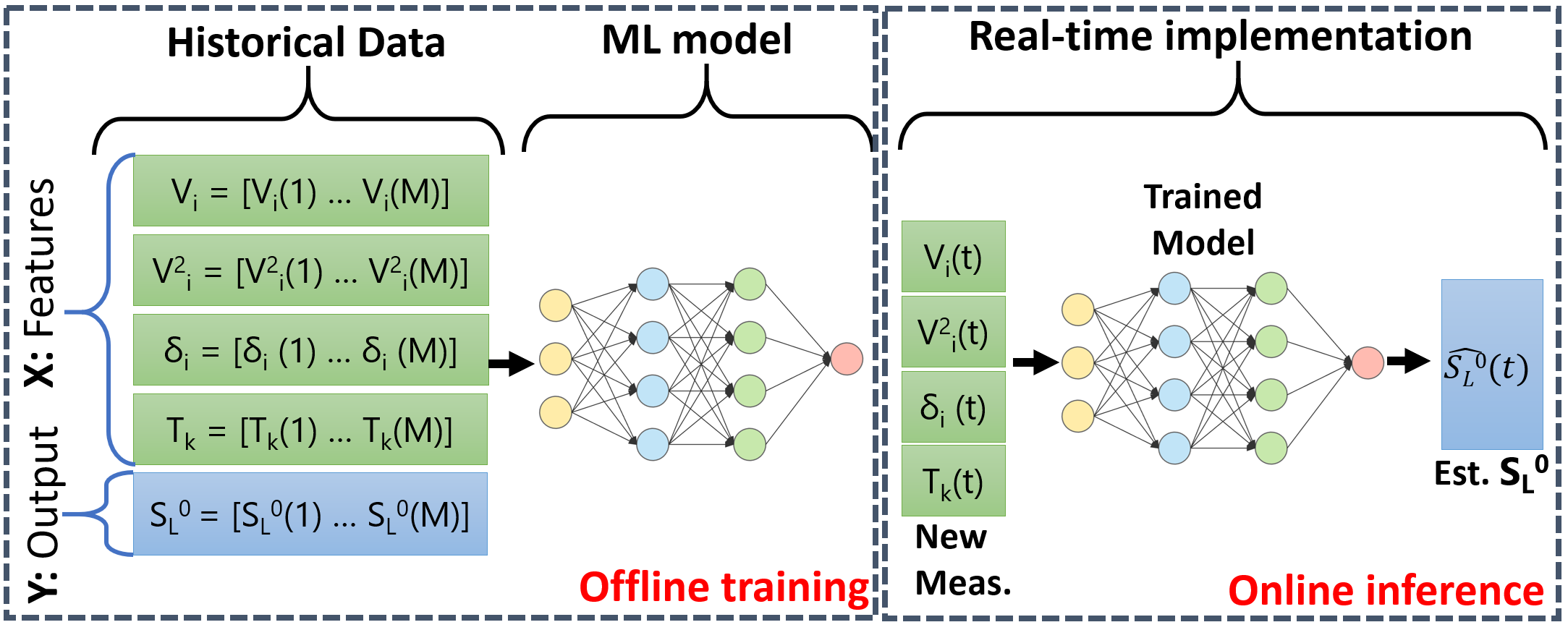}
\caption{Training and implementation phases of the ML network using historical voltage and substation power data. }
\label{fig:learning_schemetic}
\end{figure}

For the ML model, a fully connected neural network (NN) with 4 layers of 30, 20, 10, and 5 neurons are constructed as shown in Fig.~\ref{fig:neural_network}. End-users train their ML models where the inputs to the ML model are the time-series local voltage measurements ($\mathbf{V_{i}}{=}[V_{i}(1)\cdots V_{i}(M)]^{T}$), phase measurements ($\bm{\delta_{i}} = [\delta_{i}(1) \cdots \delta_{i}(M)]^{T}$), and time interval vector ($\mathbf{T_{k}} = [T_{k}(1) \cdots T_{k}(M)]$ where $T_{k}(t) = \ceil{\frac{tk}{24\times3600}} $, $k$ is the number of time intervals during a day and $t$ is in seconds). The outputs (labels) are time-series total substation feeder apparent power measurements ($\mathbf{S_{L}^{0}}$). As an additional feature, we also included the voltage square ($V_{i}^{2}$) in the inputs since \eqref{eq:VoltageVsPower} directly relates to $V_{i}^{2}$. After training the model, the new real-time local measurements can be fed to the NN in the implementation phase to estimate the substation loading $S_{L}^{0}(t)$ as illustrated in Fig.~\mbox{\ref{fig:learning_schemetic}}. 

\begin{figure}[t]
\centering
\includegraphics[scale=0.24 ]{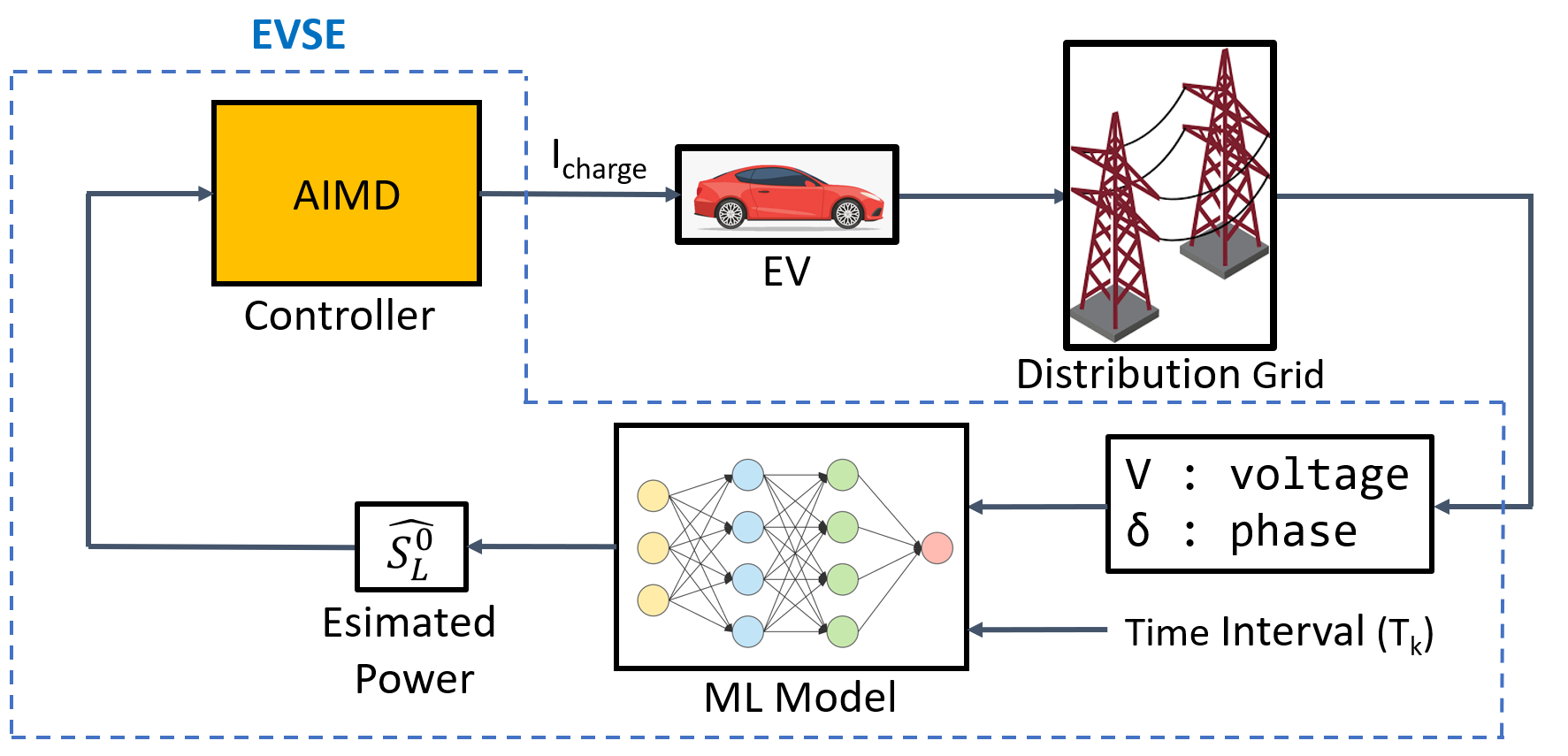}
\caption{Control structure for EV charging.}
\label{fig:control_method}\vspace{-0.5cm}
\end{figure}

The simulation time step was chosen as one second. Considering the grid dynamics and fast load fluctuations, the estimation of $S_{L}^{0}(t)$ is performed every minute by collecting $60$ samples of input features ($V_{i},V_{i}^{2},\delta_{i},T_{k}$) and averaging them. The averaged quantities are normalized and fed through the NN generating the estimated substation total demand power, i.e., $\hat{S}_{L}^{0}(t)$. The comparison result of $\hat{S}_{L}^{0}(t)$ with the rated substation capacity $S_{L(rated)}^{0}$ is used in the AIMD algorithm to decide whether there is a CE. This comparison check is performed at the same period of 60s (algorithm period) and EV charging current is either increased additively or decreased multiplicatively based on the CE decision. The additive ($\alpha$) and multiplicative ($\beta$) parameters are set to $1$ and $0.5$, respectively. The proposed control algorithm is presented in Algorithm~\ref{alg1}, and its implementation is illustrated in Fig.~\ref{fig:control_method}.

\begin{algorithm}
\caption{AIMD algorithm for EV charging network}\label{alg1}
\hrule
 \vspace{1mm}
\begin{algorithmic}[1]
 \item[\textbf{Input:}]Substation rated capacity: $S_{L_{(rated)}}^{0}$
 \item[\textbf{Input:}]Voltage meas. : $V_{i}(t)$, $V_{i}^{2}(t)$
 \item[\textbf{Input:}]Phase meas. : $\delta_{i}(t)$
 \item[\textbf{Calculate:}]Time interval : $T_{k}(t)$
 \item[\textbf{Estimate:}]  $\hat{S}_{L}^{0}(t)$ using ($V_{i},V_{i}^{2},\delta_{i},T_{k}$) thru NN
 \item[\textbf{Parameter:}] Additive parameter: $\alpha_{i}=1$
 \item[\textbf{Parameter:}] Multiplicative parameter: $\beta_{i}=0.5$
 \item[\textbf{Parameter:}] Minimum voltage threshold: $V_{min}=0.9$~pu
 \item[\textbf{Input:}]Previous charging current: $I_{i}(t)$
 \item[\textbf{Output:}] New charging current: $I_{i}(t+1)$

\WHILE{SOC $<$ 100\%}
 \IF{$\hat{S}_{L}^{0}(t) < S_{L_{(rated)}}^{0}$ \AND $V_{i}(t) > V_{min}$}
 \STATE $I_{i}(t+1)=I_{i}(t)+\alpha_{i}$
 \ELSE
 \STATE $I_{i}(t+1)=\beta_{i}\times I_{i}(t)$
 \ENDIF
 \ENDWHILE
 \vspace{1mm}
\hrule
\end{algorithmic}
\end{algorithm}
 \vspace{-3mm}
 
\section{Test System Modeling} \label{sec:system_modeling}
\begin{figure}[tb]
\vspace{-0.25cm}
  \centering
	\includegraphics[width=90mm]{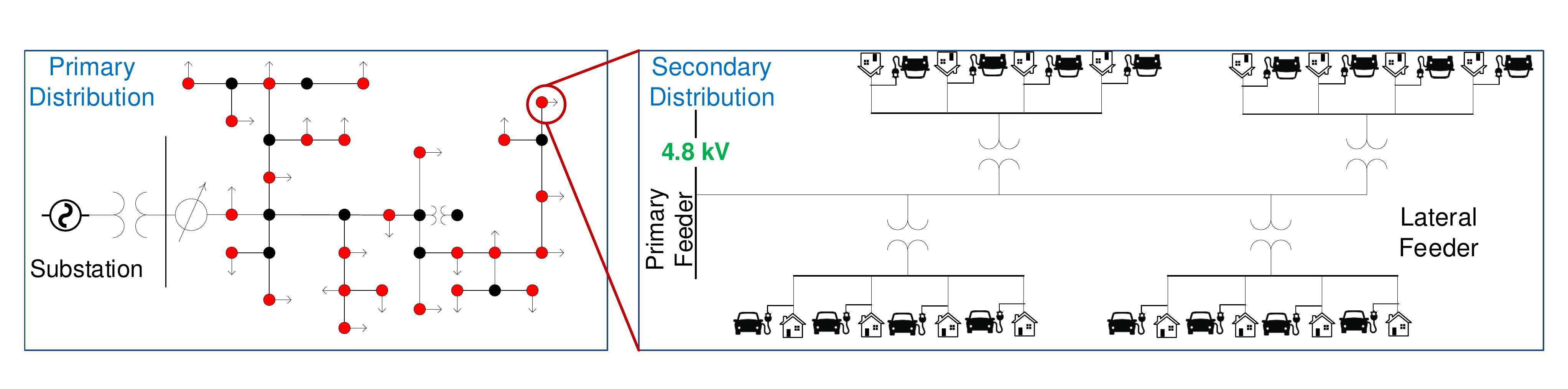}
	\caption{Primary and secondary distribution network implemented in the MATLAB model.}
    \label{fig:grid}
\end{figure}

The simulations are performed in an IEEE-37 test distribution grid illustrated in Fig.~\ref{fig:grid}. Each red circle in the figure represents a neighborhood that is connected to the primary network. There are a total of 26 neighborhoods, each of which is modeled as a secondary network following a similar procedure described in \cite{malekpour2015radial}.  Each neighborhood contains four 25~kVA transformers ($26\times4=104$ transformers in total) powering four inner nodes. The transformers step down the primary feeder voltage of 4.8~kV to a secondary voltage level of split-phase 120/240~V. Each inner node consists of 4 houses making 16 houses in a neighborhood and 416 houses in the overall distribution model. Each house is modeled as two separate nodes; one node is dedicated only to EV connection, and the other node is reserved for the uncontrollable household load. Hence, there are $416\times2 = 832$ separate nodes in the distribution network. Each end-node represents a residential customer that has a unique household power consumption profile and an EV that is modeled separately. The capacity of the substation transformer is rated at 2.5~MVA, and the grid operates slightly over 1.36~MVA at peak hours without any EV charging event, i.e. the base load.

\begin{figure}[tb]
\centering
\includegraphics[scale=0.21]{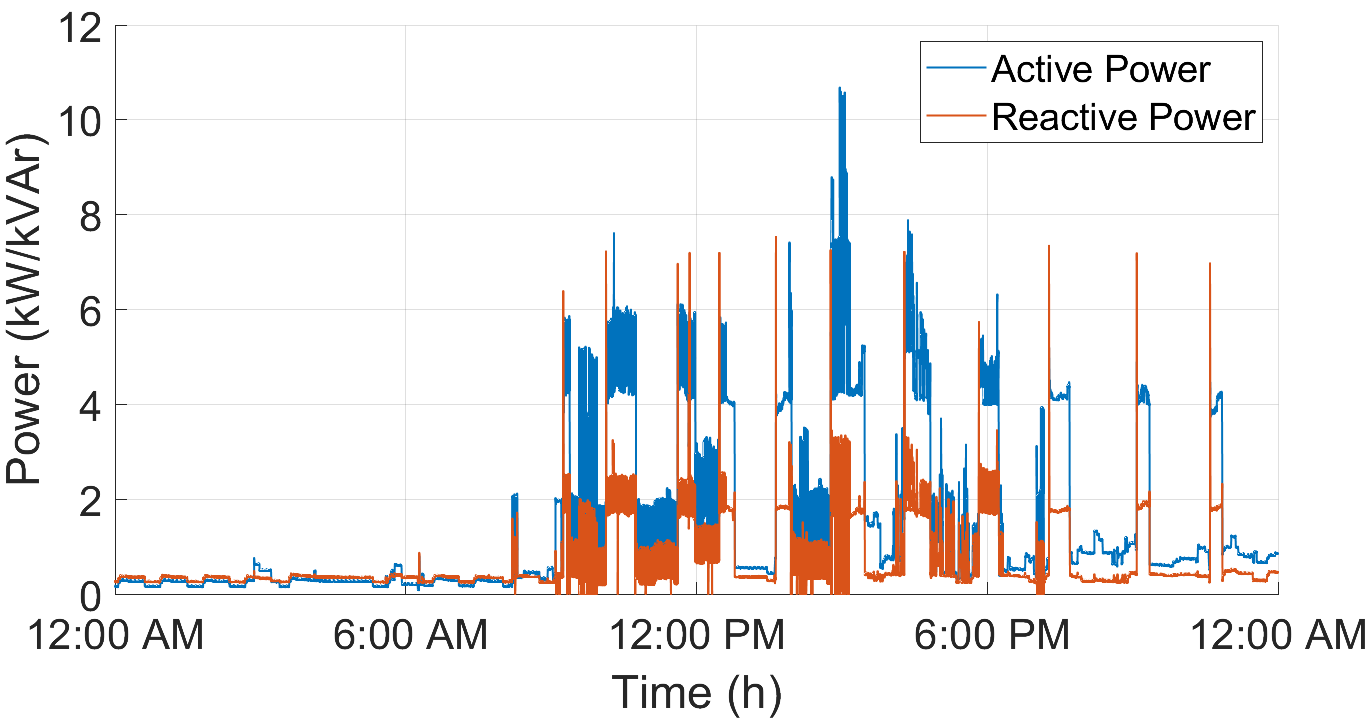}
\caption{An example of real active and reactive power profile.}
\label{fig:LoadProfiles}
\vspace{-0.5cm}
\end{figure}

These residential power consumption profiles are based on actual field measurements collected at a household located in Alabama for over 500 days using eGauge smart energy metering system. The collected data contains a one-second resolution active and reactive power consumption profile. An example profile is demonstrated in Fig.~\ref{fig:LoadProfiles}.

\section{Results and Discussion} \label{sec:results}
For the training phase, we simulated for 30 subsequent days with different load profiles for each node chosen from a set of 500+ days of collected data. EV penetration was increased by 3.3\% every day from 0 to 100, and EVs were charged at the rated power of 7.2~kW. The end-node voltage and phase measurements as well as the substation loading were saved to train a NN for each end-node. Algorithm~\ref{alg1} is implemented for each EV in a day-long simulation with a step time of 1~s at 100\% EV penetration. To better assess the performance of the \emph{proposed AIMD} method, the \emph{ideal version of AIMD} where the CE is directly sent to end-nodes is implemented. We also included the results for the \emph{No Control} case where EVs are charged at rated power without any control.

\begin{figure}[tb]
\vspace{-0.25cm}
\centering
\includegraphics[scale=0.25]{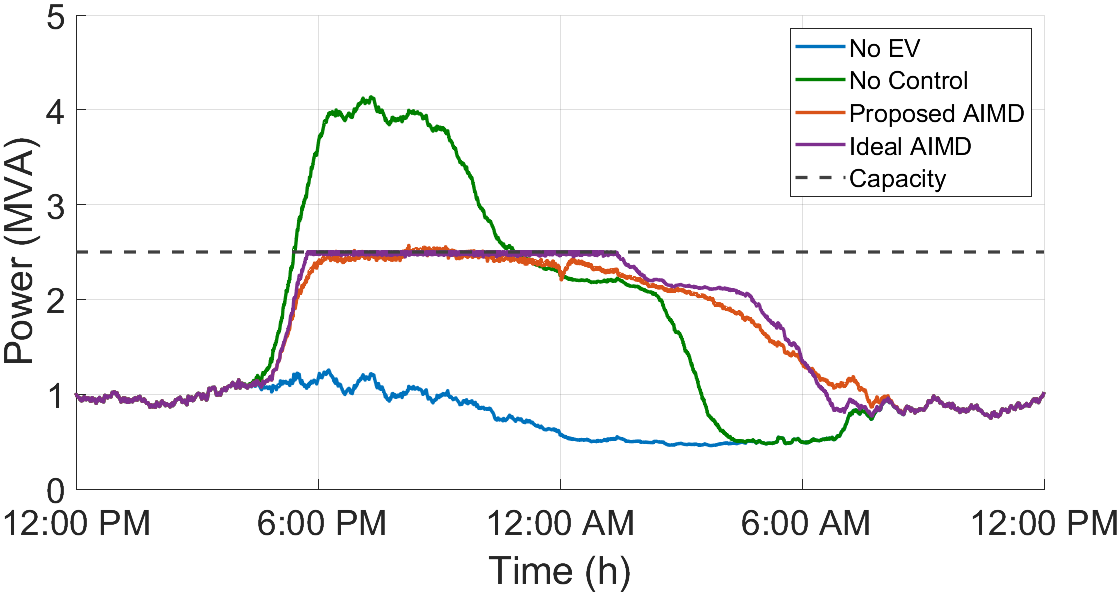}
\caption{Total substation loading for all use cases.}
\label{fig:TotalS}
\end{figure}

Fig.~\ref{fig:TotalS} shows the total substation power loading for all simulation results. \emph{No EV} and \emph{No Control} cases are included for comparison purposes. We see that the \emph{proposed AIMD} performs similar to the \emph{ideal AIMD} and manages to keep the substation loading below the capacity limit of 2.5~MVA without using any real-time communication but only local voltage measurements.

\begin{figure}[tb]
\centering
\includegraphics[scale=0.25]{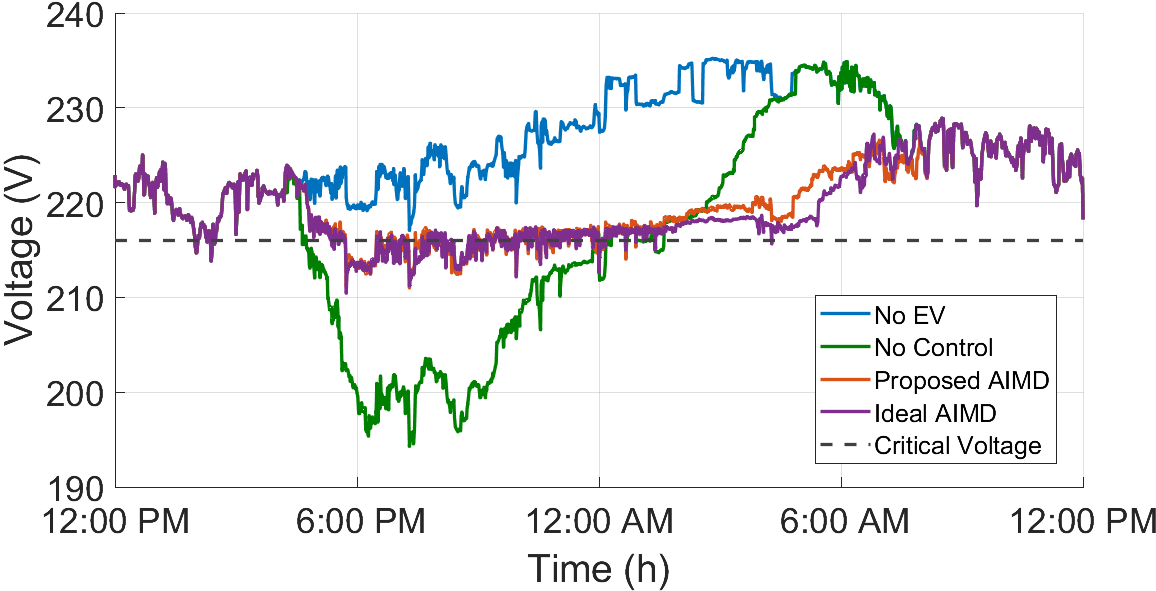}
\caption{Minimum voltage profiles for all use cases.}
\label{fig:MinVoltages}
\vspace{-0.5cm}
\end{figure}

In Fig.~\ref{fig:MinVoltages}, we plot the minimum voltage observed in the grid among all 416 nodes at a given time for all simulations. We see that the voltage goes significantly below the critical level of 216~V when EVs are not controlled. It is clear that the grid experiences this voltage drop during the substation overloading, and by preventing the overloading, the \emph{proposed AIMD} also avoided the severe voltage drop.

\begin{figure}[tb]
\vspace{-0.3cm}
\centering
\includegraphics[scale=0.25]{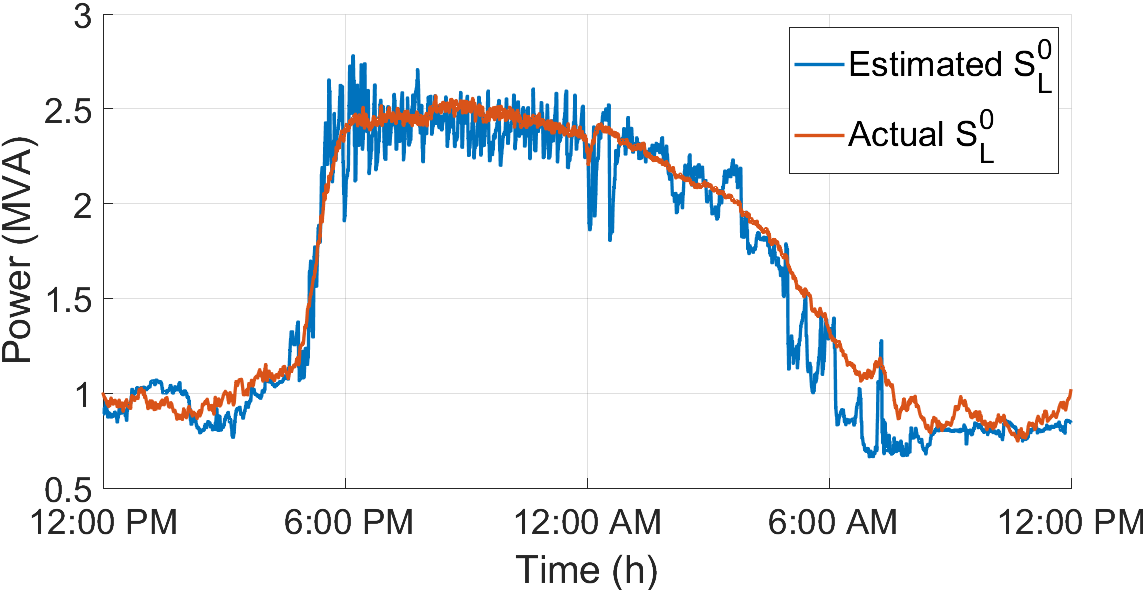}
\caption{Estimated substation loading of a random end-node.}
\label{fig:Estimated}
\end{figure}

Fig.~\ref{fig:Estimated} shows the estimated substation loading ($\hat{S}_{L}^{0}(t))$ of a randomly selected node during the simulation. The trained NN model successfully predicted the substation loading, especially during the peak time (6:00~PM-12:00~AM). Noisy predictions could be eliminated by increasing the averaging window of voltage at the expense of more delayed responses. 

\begin{figure}[tb]
\centering
\includegraphics[scale=0.23]{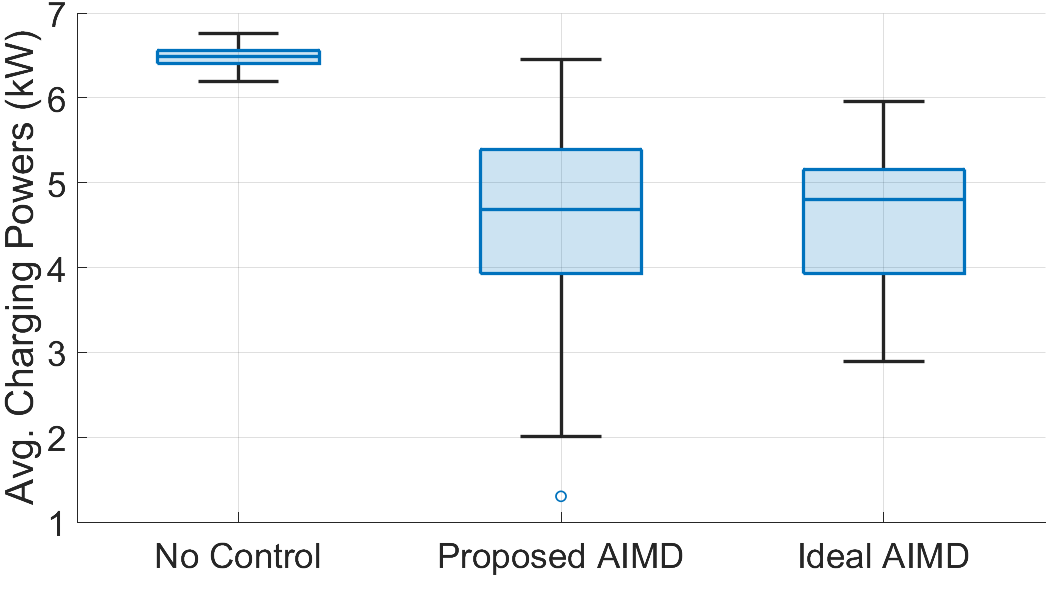}
\caption{An example of real active and reactive power profile.}
\label{fig:AvgPowers}
\end{figure}

Fig.~\ref{fig:AvgPowers} shows the average charging power distributions for the simulated cases. The \emph{proposed AIMD} resulted in an average charging power of 4.58kW, which is very close to 4.60kW of the \emph{ideal AIMD}. The standard deviations of average charging powers for the \emph{proposed AIMD} and \emph{ideal AIMD} are 1~kW and 0.71~kW, respectively. This is mainly due to the nodes receiving the true congestion signal simultaneously in the \emph{ideal AIMD} case, resulting in a fairer charging. 

Finally, Table~\ref{tab:score_table} summarizes some key performance scores for the two algorithms along with the \emph{No Control} case. Max. Substation Overload refers to how much the substation capacity is overloaded in percentage. The \emph{proposed AIMD} is slightly over the capacity (2.73\%), but this is certainly tolerable, especially for the considered overload duration (Fig.~\ref{fig:TotalS}). Moreover, this can also be avoided by setting a margin for the CE condition such that $\hat{S}_{L}^{0}(t) < S_{L_{(rated)}}^{0} - \epsilon$ for $\epsilon > 0$. Avg. SOC is the average of the final State of Charge (SOC) of EVs. The results show that the \emph{proposed AIMD} manages to satisfy more than 96\% of all EV customers' energy demand. Fairness Score is Jain's fairness index \cite{Jain1989Fairness}, which represents how fair a resource is allocated among $N$ users. One (1) means the fairest allocation (equal share per user), and $1/N$ is the least fair index. The \emph{proposed AIMD} has a very high fairness score of 0.955, slightly below the \emph{ideal AIMD} but requires no real-time communication whereas the \emph{ideal AIMD} has to receive the CE signal every second, resulting in a Number of Communication Exchange of 86400. The \emph{proposed AIMD} must receive the substation loading data only once for training and then operate on its own by solely using local voltage measurements.

Our findings show that the \emph{proposed AIMD} performs very close to its ideal counterpart in terms of grid constraints (i.e., overloading, voltage violation, etc.) and customer satisfaction (i.e., avg. charging power, SOC, fairness, etc.). It also leads to more scalable and autonomous EV integration due to its model-free and real-time communication-free features.

\begin{table}[tb] 
\caption{Performance comparison with different algorithms.}
\vspace{-2mm}
  \centering
\resizebox{\columnwidth}{!}{  
\begin{tabular}{l|lllll}
\hline
Algorithm&Max. Substation  &Avg. &Avg. &Fairness & \# of Comm.\\
&Overload (\%)&Power (kW) &SOC (\%) &Score &  Exchange\\
 \hline
 \hline
No-Control& 65.48 & 6.47 & 99.00 & 0.999 & 0  \\
Proposed AIMD& 2.73 & 4.58 & 96.04 & 0.955 & 1   \\
Ideal AIMD& 0.15 & 4.60 & 98.72 & 0.976 & 86400   \\
\hline
\end{tabular} 
}
\label{tab:score_table}
\vspace{-4   mm}
\end{table}

\vspace{-0.25cm}
\section{Conclusion}\label{sec:conclusion}
This study proposed a local, data-driven EV charging control method based on the AIMD algorithm. To eliminate the real-time communication and feedback for the CE, we proposed a method to estimate the substation loading power using local voltage measurements. We tested the proposed algorithm against the ideal AIMD and concluded that our algorithm performs very close to the ideal AIMD in peak-load management, voltage regulation, and average charging (fairness). In the future, we plan to improve the estimation algorithm further to make it robust to grid-related disturbances such as On-Load Tap Changer (OLTC) and capacitor-bank switching actions.

\bibliographystyle{IEEEtran}
\bibliography{./ref.bib}

\end{document}